\documentclass[12pt, a4wide, lualatex, english]{article}

\usepackage{newtxtext,newtxmath}

\usepackage{latexsym}
\usepackage{graphicx}
\usepackage{amsfonts} 
\usepackage{amsmath}
\usepackage{bm}
\usepackage{color}
\usepackage{tikz}
\usepackage[italicdiff]{physics}
\usepackage{version}
\usepackage{natbib}
\usepackage{setspace}
\usepackage{url}
\usepackage{comment}
\usepackage[none]{hyphenat}
\usepackage{here}
\usepackage[top=25truemm, left=25truemm,right=25truemm]{geometry}
\usepackage[labelformat=simple]{subcaption}

\usepackage{pgfplots}
\usepackage{tikz}
\usetikzlibrary{backgrounds}
\usetikzlibrary{intersections}

\def\col{\overline{c}}

\def\thetamu{\mu} 
\def\Nescape{N_{t, \mathrm{escape}}}
\def\Nstarve{N_{t, \mathrm{starve}}}
\def\Ntildeescape{\widetilde{N}_{\mathrm{escape}}}
\def\Ntildestarve{\widetilde{N}_{\mathrm{starve}}}
\def\Ntildess{\widetilde{N}_{\mathrm{ss}}}
\def\nss{n_{\mathrm{ss}}}
\def\nescape{n_{\mathrm{escape}}}

\begin{document}
\onehalfspacing
\title{\vspace{-0cm}A  Pomeranzian Growth Theory\\
 of the Great Divergence
}
\author{Shuhei Aoki\thanks{Email: shuhei.aoki@gmail.com.
I would like to thank the seminar participants at Tohoku University and Osaka University, as well as the participants at the 2022 Asian Meeting of the Econometric Society in East and South-East Asia.
I am particularly grateful to Tomoaki Kotera, Akira Momota, Daishin Yasui,
Yuta Takahashi, and Naoki Takayama
for their comments and suggestions.
This work was supported by JSPS KAKENHI Grant Numbers 
JP18H00831, JP19K01671, 23H00796, and 23K01399.
}\\
{Department of Economics, Shinshu University}
}

\maketitle
\begin{abstract}
This study constructs a growth model of the Great Divergence that formalizes \citeauthor{pomeranz2000}'s (\citeyear{pomeranz2000}) hypothesis that the relief of land constraints in Europe has caused divergence in economic growth between Europe and China since the 19th century.
The model consists of the agricultural and manufacturing sectors. The agricultural sector produces subsistence goods from land, intermediate goods from the manufacturing sector, and labor. The manufacturing sector produces goods from labor, and its productivity grows through the learning-by-doing of full-time manufacturing workers. Households make fertility decisions. 
In the model, a large exogenous positive shock in land supply causes the transition of the economy from the Malthusian state, in which all workers are engaged in agricultural production and per capita income is constant, to the non-Malthusian state, in which the share of workers engaged in agricultural production gradually decreases and per capita income grows at a roughly constant growth rate.
The quantitative predictions of the model provide several insights into the causes of the Great Divergence.

\noindent
Keywords: Industrial Revolution; Great Divergence; Malthusian economy; economic growth
  
\noindent
JEL classification codes: J13, O11, O33, O41
\end{abstract}

\section{Introduction}
The Industrial Revolution was a major economic event that brought about irreversible economic change. Since the Industrial Revolution, the world economy has transformed from stagnancy, in which per capita income was roughly constant, to growth, in which per capita income has grown steadily.\footnote{
There is some debate on whether per capita income increased before the Industrial Revolution.
For example, in the \citeauthor{maddison2001} database (see, e.g.,  \citealp{maddison2001}), the per capita income increased globally even before the Industrial Revolution.  Following \cite{clark2007} and \cite{clark2008},
this study assumes that per capita GDP was roughly constant before the Industrial Revolution (see also \citealp{broadberry-etal2015}).
}
The reasons behind the occurrence of the Industrial Revolution have been the subject of numerous studies.

\begin{figure}[htbp]
\begin{center}
\caption{GDP per capita in the pre-modern era}\label{fig:gdppc}
\vspace*{0cm}
\includegraphics[ width=.6\columnwidth]{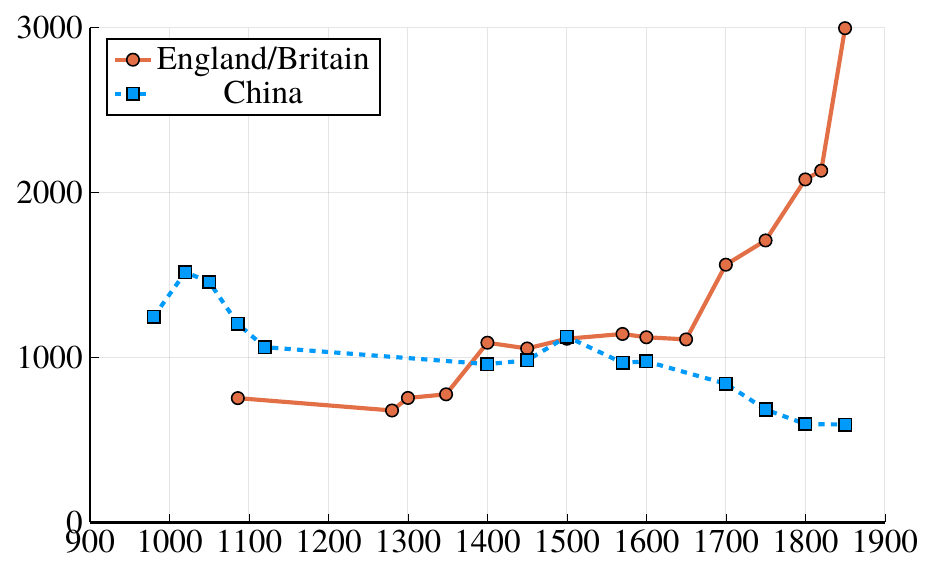}
\vspace*{-.5cm}
\end{center}
{\footnotesize Notes: Data are obtained from Table 10.02 in \cite{broadberry-etal2015}.}
\end{figure}

By considering comparative historical aspects, \cite{pomeranz2000} makes a breakthrough in this problem.
He argues that in the 18th century, China was at a similar development level to that of Europe. 
The levels of science, technology, institutions, and market efficiency in China were not far behind those in Europe. 
The per capita GDP of the Yangzi Delta, the most developed region in China, was roughly at par with that of the Netherlands, the most developed region in Europe.
This claim is supported by recent estimates of GDP per capita, as plotted in Figure \ref{fig:gdppc}. The figure shows that Britain's GDP per capita was similar to that of China before 1750, but diverged after that.
To understand why the Industrial Revolution occurred in Britain in the 18th century, researchers need to identify factors that caused what he refers to as the Great Divergence--the divergence in economic growth between Europe and China since the 19th century. These factors should have been present in Britain but absent in China.

According to \cite{pomeranz2000}, the key factors are coal and imports from the New World. 
In the pre-modern era, 
all necessities were produced using land. 
Food and material for clothing were produced through farmlands and ranching.
Building material and fuel were collected from the forests.
The expansion of farmland, ranching, and timber extraction caused deforestation, resulting in soil erosion, floods, and water shortages.
Under these ecological pressures, pre-modern societies faced limitations in their growth. 
In Britain, coal and imports of agricultural goods such as cotton, sugar, cereals, and timber from the New World, relieved land constraints. \cite{pomeranz2000}, extending \citeauthor{wrigley1988}'s (\citeyear{wrigley1988}) idea,
 argues that the relief of land constraints is the reason why the Industrial Revolution did not occur in China but in Britain.
Although \cite{pomeranz2000} stimulates interest and fosters discussion on the causes of divergence, the precise economic mechanisms underlying \citeauthor{pomeranz2000}'s thesis remain ambiguous.
For example, why did the pre-modern economy stagnate and why did the relief of land constraints generate sustained economic growth?

The purpose of this study  is to construct a formal Pomeranzian growth model of the Great Divergence and clarify the economic mechanisms behind \citeauthor{pomeranz2000}'s hypothesis.
I focus on the following two observations underlying his hypothesis.
First, in pre-modern societies, the majority of the workforce was engaged in agricultural production,
because a subsistence level of agricultural goods such as food was necessary for their survival.
To capture the point, 
I assume a hierarchical preference structure that follows Engel's law, in which households need to consume a subsistence level of agricultural goods before spending on other goods.\footnote{Hierarchical preference is an extreme limit case of the Stone-Geary preference.}
This preference structure is empirically supported and models incorporating it can predict several facts, including the declining share of agricultural employment (see \citealp{eswaran-kotwal1993}, \citealp{laitner2000}, \citealp{gollin-etal2002}, and \citealp{gollin-etal2007}).
Second, it was not until sufficient agricultural goods were provided by the relief of land constraints in Britain and labor was reallocated to the manufacturing sector that the Industrial Revolution occurred.
To capture this observation, I assume that manufacturing productivity grows through the learning-by-doing process of manufacturing workers.

\cite{matsuyama1992} constructs a model of the Industrial Revolution that incorporates the above two characteristics. His model interprets that the Industrial Revolution was caused by an increase in agricultural (labor) productivity, which reallocated labor from agriculture to manufacturing and facilitated learning-by-doing in the manufacturing sector.
However, his model does not explain why the Great Divergence occurred, that is, why agricultural (labor) productivity increased in Britain, but not in China.

This study offers an explanation for the Great Divergence by incorporating three additional features into the model.
The first additional factor introduced to explain the Great Divergence is land as an agricultural production input. Here, the relief of land constraints in the Pomeranzian hypothesis is interpreted as a sudden increase in land supply.\footnote{
Following \cite{pomeranz2000}, 
this study considers coal as a substitute for timber and assumes the following:
The use of coal affects only the supply of agricultural goods.
Although coal also affects the supply of manufacturing goods, an analysis of this case is left for future research.
}
By this assumption,
this study qualitatively describes the Great Divergence as follows: 
Suppose that an economy is in the Malthusian state, in which 
with little land supply 
and a large population, all households are forced to work in the agricultural sector to produce a subsistence level of agricultural goods.
Then, per capita income does not grow. 
The economy can produce a sufficient amount of agricultural goods with fewer households if the land supply suddenly and significantly increases because of an exogenous shock.
Then, the economy switches to the non-Malthusian state, in which 
surplus labor now works in the manufacturing sector.
The manufacturing productivity grows through learning-by-doing in the sector
and per capita income increases.
This study interprets this result as a description of the Industrial Revolution.

The second factor is the population dynamics generated from endogenous fertility choice. 
Following Malthus, Pomeranz emphasizes the dynamics in pre-modern societies in which population growth led to a shortage of resources necessary to produce agricultural goods, which in turn caused a slowdown in population growth.
To analyze these interactions, I incorporate a setting into the model 
in which households endogenously choose the number of their children and spend a portion of their labor on childcare (see \citealp{barro-becker1989}, \citealp{galor-weil2000}, and \citealp{greenwood-etal2005aer}, among others).
In these models, when the initial household income level is low, 
as household income rises, the household increases the number of children; however, as the income level exceeds a certain level, the household gradually decreases the number of children.
This setting can account for the population dynamics in pre-modern societies described above as well as the decline in fertility rate in modern societies.
To align the model with the historical fact that population growth occurred while per capita income was roughly constant in pre-modern societies, I further assume that agricultural productivity grows exogenously at a constant rate.

The third factor comprises the intermediate inputs in the agricultural production function, which are produced in the manufacturing sector.
Historically, after the Industrial Revolution, 
the share of agricultural employment 
has decreased as manufacturing products such as chemical fertilizers, tractors, and combines, which have resulted from manufacturing technological changes, have reduced the workforce required for agricultural production. To capture this effect in the model,
as in \cite{restuccia-etal2008},
this study introduces the intermediate goods produced in the manufacturing sector as inputs for agricultural production. 
It is further assumed that farmers also provide intermediate goods for the production of agricultural goods through part-time manufacturing work and that this part-time manufacturing work by farmers does not contribute to an increase in manufacturing productivity and only full-time manufacturing work increases manufacturing productivity.
These assumptions reflect the fact that in pre-modern economies, farmers themselves made fertilizers, plows, and hoes, and that the lack of division of labor impeded the development of manufacturing expertise.
These assumptions are necessary to obtain the result that a Malthusian economy is sustainable, in which all households are forced to work as farmers and per capita income remains constant.

Using this model, 
I quantitatively examine whether the economy, which was initially trapped in the Malthusian state, experienced sustained economic growth through relief of land constraints.
From his calculations in \cite{pomeranz2000}, I estimate the relief of land constraints caused by coal and imports from the New World,
corresponds to a 2.74 times increase in land supply in Britain.
Using the estimates and under other calibrated parameters, 
through a 2.74 times increase in land supply,
an economy that was initially trapped in the Malthusian state, in which per capita income is constant, switches 
into a non-Malthusian state, in which per capita income grows at a roughly constant rate.
After a 2.74 times increase in land supply, the agricultural share of employment in our model declines by the increased supply of intermediate inputs, which is quantitatively consistent with the situation in Britain.

The model developed in this study has several implications.
 First, according to \cite{pomeranz2000}, in the pre-modern era, land productivity in China was higher than that in Europe, whereas labor productivity was similar in both regions. Our model can account for this fact by assuming that China had a higher agricultural productivity level than Europe and that both economies were in the Malthusian state. The higher agricultural productivity in China was offset by its higher population level. 
Second, in the model, a sudden decline in population size, which historically occurred due to epidemics and wars, 
has an effect similar to that of increased land supply. However, I show that 
even the largest of them, caused by the Mongol Conquest of the 13th century,
is quantitatively smaller than the relief of land constraints caused by coal and imports from the New World. 
Third, our model offers another interpretation of why per capita income did not grow, whereas the population grew in the pre-modern era. This fact might be inconsistent with the prediction of endogenous growth models, in which population growth is the engine of per capita income growth. 
Our model offers the interpretation that per capita income stagnated in the pre-modern era because laborforce was allocated to the agricultural sector that engaged in subsistence production and not to productive activities, possibly factory work in the manufacturing sector that generated sustained per capita growth.

Several studies propose formal economic models that account for why and how the Industrial Revolution occurred (for surveys of these theories, see \citealp{clark2014} and \citealp{mokyr2005}). These models can be divided into two categories based on whether the Industrial Revolution is inevitable.\footnote{This classification is based on \cite{jones2001}.}
 Most models, including \cite{acemoglu-zilibotti1997}, \cite{galor-weil2000}, and
\cite{jones2001}, \cite{hansen-prescott2002}, and \cite{roy_seshadri2017},
 assume that an Industrial Revolution is inevitable and that it eventually occurs. For example, the pre-modern economy in \cite{galor-weil2000}, 
population size gradually increases. When the population increases to a certain level, the economy automatically switches from the Malthusian regime to the modern growth regime.
By contrast, in a few models, the Industrial Revolution is not inevitable.
In \cite{matsuyama1992}, an Industrial Revolution occurs because of an increase in agricultural productivity. 
\cite{lucas2002} constructs a model in which a positive exogenous shock to the rate of human capital accumulation allows the economy to transition from a stagnant to a sustained growth path. 
This study belongs to the second category of research. 
Compared with these previous studies, this study identifies the exogenous shocks that transform the economy from stagnation to growth based on economic history studies and quantitatively examines the magnitude of the shocks.

The remainder of this paper is organized as follows: Section \ref{sec:model} introduces the model. Section \ref{sec:prop} discusses the analytical properties of the proposed model.
Section \ref{sec:numerical} presents the numerical exercises to quantitatively evaluate the Pomeranzian hypothesis of the Great Divergence.
Section \ref{sec:implications} discusses the implications of the model.
Finally, Section \ref{sec:conc} concludes the paper.

\section{Model}\label{sec:model}
This section presents a two-sector growth model in which the fertility choice is endogenous.
Time is discrete, ranging from zero to infinity. 
As in \cite{lagerlof2003}, one period corresponds to 25 years. 

\subsection{Production}
The economy produces two final goods: agricultural and manufacturing. 
As in \cite{restuccia-etal2008}, agricultural goods $Y_{at}$ at time $t$ are produced from land $Z$, intermediate inputs produced in the manufacturing sector $X_t$, and labor inputs by farmers $L_{at}$:
\begin{align}
 Y_{at} =  Z^{\theta_Z} X_t^{\theta_X}\pqty{A_{at}L_{at}}^{\theta_L},\label{eq:Ya}
\end{align}
where $A_{at}$ is agricultural productivity,
$\theta_Z, \theta_X, \theta_L>0$, and 
$\theta_Z + \theta_X + \theta_L = 1$.
The time subscript is dropped from land $Z$ because land supply is constant.
Agricultural productivity $A_a$ grows at a constant rate $G_a$:
\begin{align*}
\frac{A_{a t+1}}{A_{a t}} = G_a \ge 1.
\end{align*}

Manufacturing goods $Y_{mt}$ are produced, 
using both full-time manufacturing workers $L_{mt}$ and part-time work by farmers $\mu \cdot L_{at}$, where $0 < \mu < 1$:
\begin{align}
 Y_{mt} = A_{mt} L_{mt} +  A_{mt} \mu L_{at},
\end{align}
where $A_{mt}$  is the manufacturing productivity.
Part-time work by farmers captures the fact that, in the pre-modern period, many manufacturing goods were produced through home production by farmers.\footnote{\cite{pomeranz2000} refers to the ``man plows, woman weaves'' culture in the Chinese Qing dynasty.} Moreover, even in the modern period, for example, in Japan, many farmers have worked part-time in factories.

Productivity in the manufacturing sector grows through learning-by-doing if full-time manufacturing workers are employed.
I simply assume that it grows at a constant rate $G_m$ if $L_{mt} > 0$. 
However, the productivity does not increase if the entire labor force works as farmers.
This assumption captures the observation that if farmers produce manufacturing goods by home production, as they did in the pre-modern period, expertise in production would be easily lost, as it is not exchanged efficiently between farmers.
In sum,
\begin{align*}
 \frac{A_{mt+1}}{A_{mt}} = 
\begin{cases}
 1 &\ \text{if $L_{mt}=0$,}\\
 G_{m} > 0 &\ \text{if $L_{mt} >0$.}
\end{cases}
\end{align*}
Hereafter, I denote the case where $L_{mt} = 0$ as the Malthusian state, and the other case where $L_{mt} > 0$ as the non-Malthusian state.
We can incorporate into the model the elements of endogenous growth models, wherein scientists engage in the R\&D of manufacturing goods in the non-Malthusian state, and the basic properties of the model would not change.

Both sectors are competitive, and the first-order conditions for the firm maximization problems are
 \begin{align}
  \frac{\theta_X p_{at} Y_{at}}{X_t} =&\ 1,\label{eq:Xa}\\
  \frac{\theta_L p_{at} Y_{at}}{L_{at}} + \mu A_{mt}=&\ w_t, \label{eq:wa}
 \end{align}
and if $L_{mt} > 0$, then
\begin{align}
 A_{mt} = w_t,\label{eq:wm}
\end{align}
where $p_{at}$ is the price of agricultural goods and $w_t$ is the wage for a unit of labor provided by a household.
I normalize the price of manufacturing goods to unity.
From these equations, in the non-Malthusian state, where $L_{mt} > 0$,
\begin{align}
X_t = \frac{\theta_X}{\theta_L}(1-\mu)A_{mt} L_{at} \label{eq:X}
\end{align}

In the Malthusian state, even when households do not consume manufacturing goods because their income levels are low, $L_{mt}$  can be positive because of the demand for intermediate inputs in the agricultural sector. This possibility is excluded if $\mu \ge {\theta_X}/\pqty{\theta_X + \theta_L}$,
obtained from $X_t \le A_{mt}\mu L_{at}$ and \eqref{eq:X}.
For simplicity of calculation, I hereafter assume 
\begin{align}
\mu = \frac{\theta_X}{\theta_X + \theta_L}, \text{or equivalently,   } \frac{\theta_X}{\theta_L}(1-\mu) = \mu.\label{eq:mu}
\end{align}
Then, \eqref{eq:wm} and \eqref{eq:X} are satisfied in both Malthusian and non-Malthusian states.\footnote{If \eqref{eq:wm} does not hold (i.e., $w_t > A_{mt}$) in the Malthusian state, then, from \eqref{eq:wa}, $\theta_L p_{at}Y_{at}/L_{at} + \mu A_{mt} > A_{mt}$.
From this equation and \eqref{eq:Xa}, $X_t > \theta_X/\theta_L (1-\mu)A_{mt} L_{at} = A_{mt} \mu L_{at}$, which contradicts the assumption that $X_t$ is supplied by home production in the agricultural sector.}

\subsection{Households}
The utility function of a household is
\begin{align*}
U_t = \begin{cases}
     \ln c_{at} & \text{if $c_{at} \le \col_a$,}\\
\ln \col_a +   (1-\gamma)\ln \pqty{c_{mt} + \col_m} + \gamma \ln n_t & \text{otherwise,}
    \end{cases}
\end{align*}
where $c_{at}$ and $c_{mt}$ are the consumption of agricultural and manufacturing goods, respectively, and $n_t$ is the number of children.
Two features are incorporated into the preference structure.
First, I assume a hierarchical preference for agricultural goods. As in \cite{eswaran-kotwal1993}, \cite{gollin-etal2002}, and \cite{gollin-etal2007}, households strongly prefer consuming agricultural goods, such as food, to consuming manufacturing goods or raising children when agricultural goods consumption is below a threshold level $\col_a$, but do not prefer agricultural goods any more than $\col_a$.
Second, a constant parameter $\col_m > 0$ is added in the utility from manufacturing goods. 
Following \cite{greenwood-seshadri2002} and \cite{greenwood-etal2005aer}, due to $\col_m$, households prefer raising children to consuming manufacturing goods when household income is not too low but not high enough.
As the income level exceeds a certain level, the household begins to consume  manufacturing goods and gradually decreases the number of children. 
These settings play an important role in the Great Divergence, primarily by preventing the production of manufacturing goods in the Malthusian state.

Let $y_t$ be the household income and $N_t$ be the (adult) population of the economy, which is equal to the number of households. 
Then, $y_t \equiv w_t + \pi_t$, where $\pi_t$ is the evenly distributed rent from land in the agricultural sector.
The budget constraint of a household is
\begin{align}
p_{at} c_{at} + c_{mt} + \eta n_t w_t \le y_t.\label{eq:bc1}
\end{align}
Here, I assume that a fixed time cost $\eta$ is required to raise a child, where $0<\eta<1$.

The solution to the household maximization problem is expressed as follows:
\begin{align}
 c_{mt} =&
\begin{cases}
0 & \text{if $y_t \le \displaystyle p_{at} \col_a + \frac{\gamma}{1-\gamma}\col_m$},\\
\pqty{1-\gamma}\pqty{y_t - p_{at} \col_a}  - \gamma \col_m& \text{otherwise,}
\end{cases}\nonumber\\
 n_t =&
\begin{cases}
0 & \text{if $y_t \le p_{at} \col_a$},\\
 \displaystyle \frac{1}{\eta w_t}\pqty{y_t-p_{at} \col_a} & 
\text{if $p_{at} \col_a < y_t \le \displaystyle p_{at} \col_a + \frac{\gamma}{1-\gamma}\col_m$,} \\ 
\displaystyle\frac{\gamma}{\eta w_t}\pqty{y_t-p_{at} \col_a + \col_m}
& \text{otherwise.} 
\end{cases}\label{eq:n}
\end{align}

\subsection{Equilibrium}
Market-clearing conditions are given by the following equations:
\begin{align}
\col_a N_t &= Y_{at},\label{eq:eqa}\\
c_{mt} N_t + X_t &=Y_{mt}\label{eq:eqm},\\
L_{at} + L_{mt}&=\pqty{1-\eta n_t}N_t.\label{eq:L}
\end{align}
Note that labor supply decreases in \eqref{eq:L} by $\eta n_t$ per adult because of childcare.

The endogenous variables of the model are determined once the employment shares of the agricultural and manufacturing sectors are obtained using market-clearing conditions. From \eqref{eq:Ya}, \eqref{eq:Xa},  \eqref{eq:X}, and \eqref{eq:mu},  the relative price of agricultural goods becomes
\begin{align}
 p_{at} = \frac{\thetamu^{1-\theta_X}}{\theta_X}\frac{A_{mt}^{1-\theta_X}\bqty{\ell_{at}(1-\eta n_t)N_t/Z}^{\theta_Z}}{A_{at}^{\theta_L}},\label{eq:pa}
\end{align}
where $\ell_{at} \equiv L_{at}/\bqty{(1-\eta n_t)N_t}$ is the employment share of the agricultural sector.

Finally, the population dynamics are given by
\begin{align*}
 N_{t+1} = n_t N_t.
\end{align*}
In this model, because endogenous variables are determined in a static manner, the dynamics of these variables can be computed sequentially. 

\section{Properties of  the Model}\label{sec:prop}
This section analytically investigates the properties of the model, particularly the population dynamics of the Malthusian state and the effect of a sudden increase in land supply.
Because it is difficult to completely evaluate the dynamics of the non-Malthusian state analytically, in the next section, I resort to a numerical analysis.

\subsection{Malthusian state}\label{sec:malthus}
First, I analyze the equilibrium in which the economy is in the Malthusian state, where manufacturing productivity is constant at $A_{mt} = A_{m}$.  
I further restrict the case wherein households can raise children and $n_t > 0$. The conditions for this case are as follows: 
\begin{align*}
p_{at}\col_a < y_t \le p_{at} \col_a  + \frac{\gamma}{1-\gamma}\col_m.
\end{align*}
Let $\Nescape$ be the population size
where $y_t = p_{at} \col_a  + \frac{\gamma}{1-\gamma}\col_m$
and $\Nstarve$ be the population size where $y_t = p_{at} \col_a$.
The above inequalities can then be rewritten as
\begin{align*}
\Nescape \le N_t < \Nstarve.
\end{align*}
Here, $\Nescape$ is the population size that divides the Malthusian state from the non-Malthusian state.
To make the analysis interesting, I assume that  
$\Nescape$ is strictly greater than zero so that  non-Malthusian state exists.
If the population is smaller than $\Nescape$, then 
household income becomes sufficiently high for households to demand manufacturing goods for consumption; hence, the economy escapes into the non-Malthusian state.
In contrast, $\Nstarve$ is the population size, where households can only feed themselves but cannot afford to raise their children. If the population size exceeds $\Nstarve$, then the population size in the next generation $N_{t+1}$ becomes zero.

Then, I analyze the population dynamics in Malthusian state.
Let $A_{a0}$ be initial agricultural productivity. 
Let $\nss$ be the steady-state population growth rate in the Malthusian state, which is equal to
\begin{align}
\nss \equiv G_a^{\frac{\theta_L}{\theta_Z}}.\label{eq:nMalthus}
\end{align}
I define the detrended population as $\widetilde{N}_t \equiv N_t/\nss^{t}$.
From \eqref{eq:n}, the dynamics of the detrended population between times $t$ and $t+1$ is derived as follows: 
\begin{align}
 \widetilde{N}_{t+1} = \frac{n_t}{\nss}\widetilde{N}_t,\label{eq:Nnext}
\end{align}
where, given $\widetilde{N}_t$, $n_t$ is determined by the following equation:
\begin{align}
 n_t =& \frac{1}{\eta}\bqty{1
- \frac{\col_a}{\mu^{\theta_X}}\frac{\pqty{\pqty{1-\eta n_t}\widetilde{N}_t/Z}^{\theta_Z}}{A_{m}^{\theta_X} A_{a0}^{\theta_L}}
}.\label{eq:ndyn}
\end{align}

\begin{figure}[htbp]
\begin{center}
\caption{Dynamics of detrended population $\widetilde{N}_t$ in the Malthusian state}\label{fig:Malthus}
\vspace*{0cm}
\includegraphics[ width=.6\columnwidth]{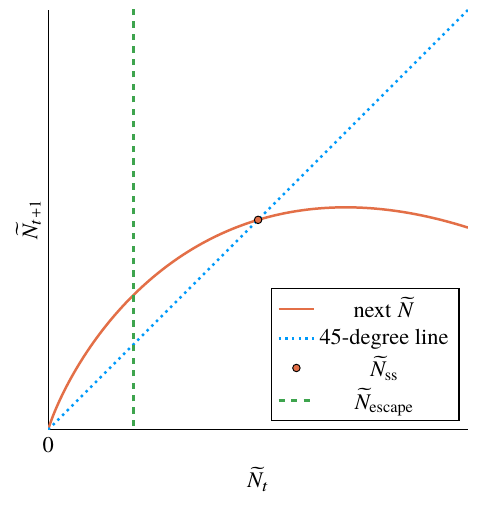}
\vspace*{-.5cm}
\end{center}

\end{figure}

The locus of \eqref{eq:Nnext} is illustrated in Figure~\ref{fig:Malthus} (see the ``next $\widetilde{N}$'' locus in the figure). 
The intersection of the locus of \eqref{eq:Nnext} and the 45-degree line is defined as $\Ntildess$:
\begin{align}
 \Ntildess = \bqty{\frac{\thetamu^{\theta_X}}{\col_a}\pqty{1-\eta \nss}^{1-\theta_Z} Z^{\theta_Z}A_{m}^{\theta_X}A_{a0}^{\theta_L}}^{\frac{1}{\theta_Z}}.\label{eq:Nss}
\end{align}
Note that the population growth rate when the detrended population level is $\Ntildess$ becomes $\nss$ given in \eqref{eq:nMalthus}.
I also add to Figure~\ref{fig:Malthus} the line for the detrended version of $\Nescape$:
\begin{align}
\Ntildeescape = 
\bqty{\pqty{\pqty{1-\theta_Z\eta \nescape} - \frac{\theta_X}{\thetamu}\frac{\gamma}{1-\gamma}\frac{\col_m}{A_{m}}}\frac{\thetamu^{\theta_X}}{\col_a }\frac{Z^{\theta_Z}A_{m}^{\theta_X}A_{a0}^{\theta_L}}{\pqty{1-\eta \nescape}^{\theta_Z}}}^{\frac{1}{\theta_Z}}.\label{eq:Ntildeescape}
\end{align}
$\nescape$ is jointly determined by $\Ntildeescape$ in Equations \eqref{eq:ndyn} and \eqref{eq:Ntildeescape}.
Note that $\Ntildess$ and $\Ntildeescape$ do not depend on time $t$ variables.
In this study, I assume that the following inequalities hold in the Malthusian state:
\begin{align}
\frac{\Ntildeescape}{\Ntildess} = \pqty{\pqty{1-\theta_Z\eta\nescape} - \frac{\theta_X}{\mu}\frac{\gamma}{1-\gamma}\frac{\col_m}{A_{m}}}\frac{1}{\pqty{1-\eta \nss}^{1-\theta_Z}\pqty{1-\eta\nescape}^{\theta_Z}} < 1,\label{eq:Nratio}
\end{align}
I do not add to Figure~\ref{fig:Malthus} the line for the detrended version of $\Nstarve$, $\Ntildestarve$, because $\Ntildestarve$ is quantitatively significantly larger than $\Ntildeescape$ and $\Ntildess$.\footnote{
Under the parameter values used in Section \ref{sec:numerical}, $\Ntildestarve \approx 3.54\times \Ntildess$.
}

The Malthusian dynamics of the model under assumption \eqref{eq:Nratio} can be analyzed from Figure~\ref{fig:Malthus}. 
Suppose the economy starts from $\widetilde{N}_0 > \Ntildeescape$.
Then, if $\widetilde{N}_0$ is not too large, the next $\widetilde{N}$ given on the right-hand side of \eqref{eq:Nnext} is greater than $\Ntildeescape$,
the detrended population converges to $\Ntildess$,
and the economy is permanently trapped in the Malthusian state.

There are several notable features of the model in the Malthusian state.
First, if $\widetilde{N}_0$ is too large, the economy escapes the Malthusian state in the next period.
In this case, the land supply relative to the population size is limited, and sufficient food is not produced. This reduces the fertility rate, resulting in a detrended population size of less than $\Ntildeescape$.
Second, if $G_a^{\theta_L/\theta_Z}$ or $A_{m}$ is too large, \eqref{eq:Nratio} is violated and the Malthusian state does not exist. 
Household behavior then shifts from raising more children to consuming manufacturing goods, causing an escape from the Malthusian state.

\subsection{Great Divergence}\label{sec:gd}
\cite{pomeranz2000} argues that coal and imports from the New World relieved land constraints and further led to the Industrial Revolution in Britain. In this study, the relief of land constraints is modelled as a sudden increase in $Z$. A sudden jump in $Z$ increases $\Nescape$. If the population $N_t$ in the economy becomes less than the new $\Nescape$, the economy can escape into the non-Malthusian state, wherein manufacturing productivity grows.

An increase in $Z$ must occur suddenly. Suppose that $Z$ increases gradually. This is accompanied by a population increase, which necessitates more labor force to be allocated to the agricultural sector to supply food to the growing population. Consequently, the economy could become trapped in the Malthusian state. This mechanism is what \cite{malthus2018} ``compared the relative speed of the growth of productive capacity and of population to a contest between a tortoise and a hare'' (\citealp{wrigley2018}). 

\subsection{Non-Malthusian state}
The economy moves into the non-Malthusian state if the population $N_t$ is strictly less than $\Nescape$. 
By rewriting \eqref{eq:eqa}, we obtain the following equation that
offers an interpretation of  the sustainability of the non-Malthusian state:
\begin{align*}
 \ell_{at} \equiv \frac{L_{at}}{N_t}= \bqty{\frac{\col_a}{\mu^{\theta_X}}  \frac{N_t^{\theta_Z}}{Z^\theta_Z A_{mt}^{\theta_X} A_{at}^{\theta_L}}
\frac{1}{\pqty{1-\eta n_t}^{1-\theta_Z}}
}^{\frac{1}{1-\theta_Z}}.
\end{align*}
Suppose that 
the population growth rate $n_t$ either decreases or does not increase significantly
and satisfies the following conditions:
\begin{align*}
 n_t < \bqty{G_m^{\theta_X}G_a^{\theta_L}}^{\frac{1}{\theta_Z}}.
\end{align*}
Then, agricultural share $\ell_{at}$ continues to decrease.
That is, if productivity growth rates are sufficiently higher than the population growth rate, the economy remains in a non-Malthusian state and per capita income continues to grow. The results also show that Malthus' tale of a tortoise and a hare applies.

The above results crucially depend on the population growth rate $n_t$. Because it is difficult to analytically investigate the dynamics of the population growth rate $n_t$, I employ numerical techniques in the subsequent section to analyze the economy in the non-Malthusian state.

\section{Numerical Exercises}\label{sec:numerical}
This section describes the numerical simulation of the model.
The purpose of this section is twofold. The first is to quantitatively examine the Pomeranzian hypothesis that 
the relief of land constraints caused by the use of coal and imports from the New World brought about the Great Divergence.
Another purpose is to numerically analyze the dynamics of the economy in the non-Malthusian state, which are difficult to explore analytically.

\subsection{Calibration}
The calibration procedure for parameter values is divided into three parts. The first focuses on the parameters of agricultural production function $\theta_Z, \theta_X$, and $\theta_L=1-\theta_Z-\theta_X$. For this, I use cost share data for the agriculture sector in the U.S. in 1960 (\citealp{capalbo-vo2015}, p. 106).
For the agricultural production function, \cite{mundlak2001} argues, ``...the elasticity of labor never exceeds 0.5, and in most cases it varies in the range of 0.25 to 0.45.’’ 
In contrast, \cite{acemoglu_restrepo2019} argue from the data in \cite{budd1960} that the income share of labor in the agricultural sector,
which should be lower than the cost share of labor,
 was 0.17 in 1910.
The value of $\theta_L$ in this study is between those reported in these studies.
\cite{mundlak2001} also writes, ``...it is meaningful to look at the sum of labor and land elasticities, and this sum is fluctuating around 0.5.'' 
$\theta_L + \theta_X$, which is equal to 0.4, is around 0.5.
This study uses these values, particularly the upper bound for $\theta_X$, to ensure that the population growth rate in the non-Malthusian state is consistent with the facts on population growth.\footnote{\label{fn}If otherwise, a lower $\theta_X$ is adopted, as $A_{mt}$ increases in the non-Malthusian state, $p_{at}$ rapidly increases  (see \eqref{eq:pa}).
Subsequently, the population growth rate, $n_t$ fluctuates (see \eqref{eq:n}).
}

The second part deals with the agricultural and manufacturing productivity growth rates.
The agricultural productivity growth rate $G_a$ is chosen such that the steady-state population growth rate in the Malthusian state, given by \eqref{eq:nMalthus}, is close to the actual population growth rate in the pre-modern era in Britain and China.
The average annual population growth rate in Britain between 1500 and 1700 was 0.43\% (see Appendix 5.3 in \citealp{broadberry-etal2015}), whereas that in China between 1480 and 1740 was 0.27\% (see Figure 2. A in \citealp{broadberry-etal2018}).
I set $G_a$ such that the annual population growth rate is 0.35\% in the Malthusian steady state.
As in \cite{lagerlof2003}, I assume that one period corresponds to 25 years.\footnote{There are some variations on the number of years for one period in quantitative models with fertility.  One period corresponds to 20 years in \cite{lagerlof2006}, while it corresponds to 35 years in \cite{hansen-prescott2002}.}
The manufacturing productivity growth rate in the non-Malthusian state is attested by the fact that the annual growth rate of output in the U.S. during the 20th century was about 2\% on average.

The third part relates to fertility choices.
In \cite{greenwood-etal2005aer}, they choose 0.17 for the utility parameter for the number of children, which is in our model $\gamma$.\footnote{
Indeed, 0.17 is the number allocated to $1-\phi$ in \cite{greenwood-etal2005aer}. 
}
\cite{lagerlof2006} chooses 0.225 for the same parameter. I set $\gamma=0.20$ to obtain the intermediate value. 
I set $\eta = \gamma/1.02$, resulting in 
the long-term annual net population growth rate in the non-Malthusian state of 0.08\%. I set $\eta$ to a value greater than $\gamma$ such that $n_t$ during the transition process in the non-Malthusian state does not fluctuate (see footnote \ref{fn}).
To simulate the dynamics, the initial population size $N_0$ must be determined. I assume that the economy begins in the Malthusian steady state, and set $N_0$ to the value of $\Ntildess$ calculated by \eqref{eq:Nss} using other parameters.

\begin{table}[htbp]
\begin{center}
\caption{Calibration of the model}\label{tab:const_params}
 \begin{tabular}{lll}
\hline
\hline
$\theta_Z$ & Share of rents from land in the agr. sector  & 0.16 \\
$\theta_X$ & Share of intermediate inputs in the agr. sector & 0.60 \\
$\theta_L$ & Share of labor inputs in the agr. sector & 0.24 \\
$G_a$ & Agricultural growth rate & $(1+0.35\%)^{\frac{\theta_Z}{\theta_L}\cdot 25}$\\
$G_m$ & Manufacturing growth rate & $(1+2.0\%)^{25}$\\
$\gamma$ & Utility parameter on the number of children & 0.20\\
$\eta$ & Child care cost & $0.20/1.02$ \\
$N_{0}$ & Initial population size & $\Ntildess$ \\
$A_{a0}$ & Initial agr. productivity & 1.0\\
$A_{m0}$ & Initial man. productivity & 1.0\\
$\col_a$ & Subsistence consumption for agr. goods & $0.25$\\
$\col_m$ &Taste parameter on man. goods & 1.35\\
\hline
\end{tabular}
\end{center}
\end{table}

The remaining parameters are either normalized to unity or 
pinned by the following restrictions.
First, $A_{a0}$, $A_{m0}$ and the initial $Z$ explained in detail in the next section are normalized to unity.
Second, the ratio of $\col_m$ to $A_{m0}$ determines $\Ntildeescape/\Ntildess$, given in \eqref{eq:Nratio}. The lower  $\col_m$ is compared to $A_{m0}$, the higher is $\Ntildeescape/\Ntildess$. If $\Ntildeescape \ge \Ntildess$, which is rewritten as 
$ \col_m/A_{m0} \lesssim  0.86$,
in the long run, the economy necessarily escapes into the non-Malthusian state. By contrast, if $\col_m/A_{m0}$ is large, it becomes more difficult for the economy to escape the Malthusian state owing to the sudden increase in land supply. 
Because ${\col_m}/{A_{m0}}$ is difficult to measure from the data,
I set ${\col_m}/{A_{m0}} = 1.35$, which results in  $\Ntildeescape/\Ntildess \approx 0.41$. 
Second, the ratio $\col_m/\col_a$ affects the change in fertility through \eqref{eq:n}. 
$\col_m/\col_a$ needs to be set
so that the dynamics of fertility depict the well-known inverted U-shaped curve in the non-Malthusian state.
In \cite{greenwood-seshadri2002}, who adopt a similar preference structure, 
$\col_m/\col_a = 1.35/0.25 = 5.4$.\footnote{In \cite{greenwood-seshadri2002}, their $a$ corresponds to $\col_a$ in my model, and their $c$ corresponds to $\col_m$. Coincidentally, their value of $c$ is similar to that of $\col_m$ in this model.
}  Following \cite{greenwood-seshadri2002}, I choose $\col_a = \col_m/5.4=0.25$ (therefore, I coincidentally set the same numbers  for $\col_a$ and $\col_m$).

\subsection{Magnitude of abolishing the land constraint}\label{sec:land}
\begin{table}[htbp]
\begin{center}
\caption{Land supply $Z$}\label{tab:land}
 \begin{tabular}{lcc}
\hline
\hline
  &  Before the  Great Divergence & After the Great Divergence  \\
\hline
Economy~1 &  1.0 & 2.74 \\
Economy~2 &  1.0 & 1.0 \\
\hline
\end{tabular}
\end{center}
\end{table}

As explained in Section \ref{sec:gd}, I model the relief of land constraints, which \citeauthor{pomeranz2000} argues was the cause of the Great Divergence and the Industrial Revolution in Britain, as a sudden increase in $Z$. 
I normalize the land supply $Z$ before the Great Divergence to unity.
I consider two cases for the periods after the Great Divergence, which I assume to occur at $t=10$: in the first case, after the Great Divergence, land supply increases, and land supply in the second case is kept constant as before. I refer to the economy in the first case as Economy~1 and in the second case as Economy~2. 
Economy~1 corresponds to Britain and Europe, whereas Economy~2 corresponds to  China.

Subsequently, we quantify the extent to which land supply increased after the Great Divergence in Economy~1.
According to \cite{pomeranz2000},
\begin{quotation}
\dots [R]aising enough sheep to replace the yarn made with Britain's New World cotton imports by would have required staggering quantities of land: almost 9,000,000 acres in 1815, using ratios from model farms, and over 23,000,000 acres in 1830. This final figure surpasses Britain's total crop and pasture land combined. It also surpasses Anthony Wrigley's estimate that matching the annual energy output of Britain's coal industry circa 1815 would have required that the country magically receive 15,000,000 additional acres of forest. If we add cotton, sugar, and timber circa 1830, we have somewhere between 25,000,000 and 30,000,000 ghost acres, exceeding even the contribution of coal by a healthy margin. (p. 276) 
\end{quotation}
Based on this calculation, I set the land supply $Z$ after the relief of land constraints to
\begin{align*}
1 + \frac{15,000,000\ \mathrm{acres} + 25,000,000\ \mathrm{acres}}{23,000,000\ \mathrm{acres}} \approx 2.74.
\end{align*}
These values are summarized in Table~\ref{tab:land}.

\subsection{Quantitative results}

\begin{figure}[htbp]
\caption{Dynamics of the economy}\label{fig:numerical}
\begin{subfigure}[b]{.5\linewidth}
\centering
\includegraphics[width=1\columnwidth]{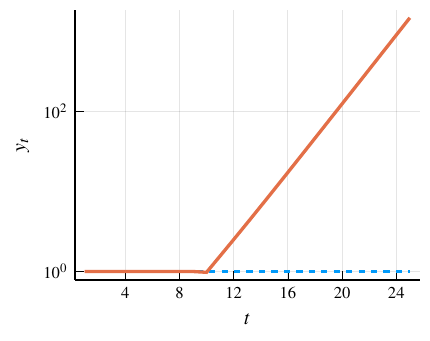}
\vspace*{-1.0cm}
\subcaption{Per capita income}\label{fig:y}
\end{subfigure}
\begin{subfigure}[b]{.5\linewidth}
\centering
\includegraphics[width=1\columnwidth]{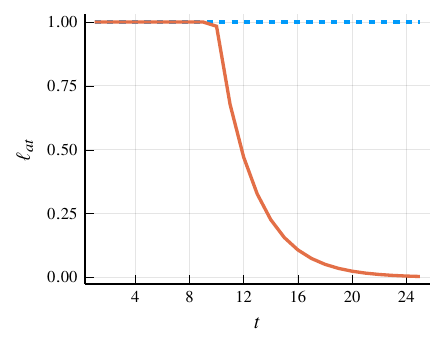}
\vspace*{-1.0cm}
\subcaption{Share of agricultural employment}\label{fig:la}
\end{subfigure}
\begin{subfigure}[b]{.5\linewidth}
\centering
\includegraphics[width=1\columnwidth]{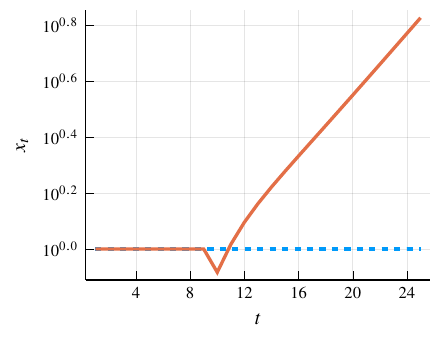}
\vspace*{-1.0cm}
\subcaption{Per capita intermediate inputs for agriculture}\label{fig:x}
\end{subfigure}
\begin{subfigure}[b]{.5\linewidth}
\centering
\includegraphics[width=1\columnwidth]{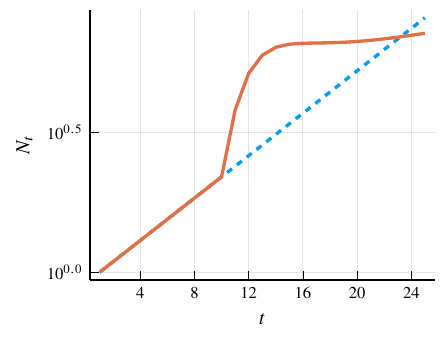}
\vspace*{-1.0cm}
\subcaption{Population}\label{fig:N}
\end{subfigure}
{\footnotesize Notes:
The solid lines in the figures are those of the economy wherein the land supply suddenly increases from $Z=1.0$ to $Z=2.74$ at $t=10$ (Economy~1).
The dashed lines are those of the economy that does not experience the increase in land supply (Economy~2). One period is equivalent to 25 years; that is, four periods are equivalent to one century. I normalize the initial values of per capita income, per capita intermediate inputs for agriculture, and population to be unity.
}
\end{figure}

This section describes the dynamics of the model. In Economy~1, land supply suddenly increases from $Z=1.0$ to $Z=2.74$ at $t=10$. In Economy~2, the land supply is kept constant.
$t=10$ corresponds to 1725 or 1750 in the real world, $t=12$ corresponds to 1775 or 1800, and $t=20$ corresponds to 1975 or 2000.

The results of the numerical exercises are shown in Figure~\ref{fig:numerical}.
In the figures, the solid lines are the paths of Economy~1 and the dashed lines are the paths of Economy~2.
Figure~\ref{fig:y} shows that 
per capita income in Economy~1 grows at an approximately constant rate after $t = 10$ (the annual growth rate after $t=10$ is approximately 1.97\%).
This is because Economy~1 switches from the Malthusian state to the non-Malthusian state at $t=10$.
Conversely, Economy~2 remains in the Malthusian state, and consequently,
per capita income in Economy~2 is constant.

Figure~\ref{fig:la} illustrates that 
in Economy~1, the share of agricultural employment significantly decreases after $t=10$. Quantitatively, the agricultural employment share $\ell_{at}$ in Economy~1 is 47\% at $t=12$ and 2.5\% at $t=20$.
Employment share of Britain's primary sector 
which excludes mining, was 48.0\% in 1759 and 41.7\% in 1801-1803
(\citealp{crafts1985british}, pp. 11-15; \citealp{shaw-taylor-wrigley2014}, p. 56).
Around 2000, the employment share of the agricultural sector was less than 5\% in most developed countries.
Although our model and assumptions are simple, the change in the agricultural employment share roughly matches the facts experienced in many developed countries.
Alongside the structural transformation, in Economy~1, the per capita intermediate inputs used for agricultural production $x_t$ grow constantly (see Figure~\ref{fig:x}).\footnote{$x_t$ in Economy~1 decreases at $t=10$, because enough level of the agricultural good to feed the population is produced with a smaller amount of $x_t$ by the sudden land supply increase.
}

Finally, Figure~\ref{fig:N} plots the populations of the economies. Because the growth rate of agricultural productivity is positive, the population grows in both economies before $t=10$.
In Economy~1, the population increases rapidly after $t = 10$ because the constraint on food supply is relieved. After this steep increase, the growth rate declines and almost stops, because households prefer manufacturing goods to raising children. 
In contrast, the population of Economy~2 grows steadily. Consequently, the population in Economy~2 exceeds that in Economy~1 at $t=24$.

\section{Implications of the Model}\label{sec:implications}
This section interprets several historical facts addressed in previous studies through the lens of the model. 
\subsection{Higher agricultural productivity in China}
\cite{pomeranz2000} summarizes several historical studies and concludes, ``agricultural labor productivity in the Yangzi Delta was still within 10 percent of English levels even near 1820, while its land productivity was several times higher---so that its total factor productivity in agriculture far exceeded that of any European locale'' (p. xvi). Our model is consistent with these findings. 

To demonstrate this property, suppose that both Britain and China are in a Malthusian steady state, whose populations are given by \eqref{eq:Nss}.  All exogenous variables are equal in both countries, except that the agricultural productivity level in China, $A_{a0}^C$ is higher than that in Britain, $A_{a0}^B$ (we denote variables with the superscript $B$ or $C$ as those in Britain or China, respectively).
Using \eqref{eq:Ya}, \eqref{eq:Nss}, and $L_{at} = (1-\eta \nss)N_t$ in the Malthusian state, we can easily show that, in the steady state, labor productivity is the same, whereas land productivity in Britain is lower than that in China:
\begin{align*}
 \frac{Y_{at}^B}{L_{at}^B} =  \frac{Y_{at}^C}{L_{at}^C},\ \ 
\text{and }\ \ 
 \frac{Y_{at}^B}{Z^B} <  \frac{Y_{at}^C}{Z^C}.
\end{align*}
Interestingly, even if the agricultural productivity level in China is higher than that in Britain, it does not contribute to escaping the stagnant Malthusian state in our model.

\subsection{Epidemics and wars}
In this model, if the population of the economy suddenly declines in Malthusian state and becomes less than $\Nescape$, the economy moves into the non-Malthusian state. In the pre-modern era, a sudden population decline occurred because of epidemics and wars. A representative example of such an epidemic is the Black Death of the 14th Century.
According to \cite{griffin2018},
``[I]n England, somewhere between a third and a half of the population are estimated to have died'' from the black death in the fourteenth century (\citealp{griffin2018}, p. 30; see also \citealp{hinde2003}, pp. 44-47).
For wars, \cite{pinker2011} collects data on the ``100 worst wars and atrocities in human history'' from several studies (see \citealp{pinker2011}, Figure~5-3). 
In his data, the worst in terms of the death toll relative to the world population 
is the Mongol Conquest of the 13th century. One of the countries most damaged by the invasion is Hungary, which lost at most 50\% of its population due to the invasion (\citealp{berend2001}, pp. 36-37).

Compared with these numbers, the relief of land constraints by coal and New World imports is quantitatively large. In the model, a 2.74-time increase in land supply, calculated in Section \ref{sec:land} from the estimates in \cite{pomeranz2000}, increases $\Ntildeescape$ by 2.74 times. This increase in land supply has the same effect as when approximately two-thirds of the population suddenly disappear, which is the upper bound of the epidemic and war estimates.
Considering that production expertise could easily be lost in the chaos of epidemics and wars, the increased supply of land by coal and New World imports is quantitatively appealing as a hypothesis of the Great Divergence.

\subsection{Prediction of endogenous growth models in the pre-modern era}
In the pre-modern era, while the population grew, per capita income remained constant. This fact is seemingly inconsistent with the prediction of the endogenous growth theory that population growth is the engine of economic growth.
One justification for this is that, in the pre-modern era, human capital accumulation was too slow because the population size was small, resulting in negligible per capita income growth, 
as explained by the unified growth model of \cite{galor-weil2000}.
Our model provides an alternative interpretation for this inconsistency.
Suppose that R\&D activities in the manufacturing sector are the engines of modern economic growth.  In my model, even if the population grows, per capita income stagnates in the pre-modern era because all labor forces are allocated to agricultural production or raising children
rather than in the manufacturing production.

\section{Conclusion}\label{sec:conc}
This study constructs a growth model that formalizes the idea of \cite{pomeranz2000} and other economic historians that the relief of land constraints by coal and imports from the New World brought about the Industrial Revolution. In this model, all workers initially engage in agricultural production to produce goods necessary for subsistence. In such an economy, per capita income is maintained at a subsistence level.
An exogenous positive shock in land supply allows the production of adequate subsistence goods with limited labor and lifts some workers from subsistence production to full-time manufacturing production. Manufacturing productivity has grown because of learning-by-doing of full-time workers in the manufacturing sector. This results in sustained growth in per capita income.
The magnitude of the increase in land supply was quantified and quantitative exercises were conducted. The predictions of the model are consistent with these facts. 
The model presented in this study seems to be a plausible candidate for the theory of the Great Divergence.

\bibliographystyle{econ}
\bibliography{econ}

\appendix

\end{document}